\documentclass[aps,prl,twocolumn,superscriptaddress,nofootinbib]{revtex4-2}
\usepackage{amsmath,amssymb,graphicx,bm}
\usepackage[T1]{fontenc}
\usepackage{physics}
\usepackage{hyperref}
\usepackage{dcolumn}
\usepackage{booktabs}
\usepackage{multirow}
\usepackage{slashed}
\usepackage[dvipsnames]{xcolor}
\usepackage[caption=false]{subfig}
\usepackage{float}
\usepackage[normalem]{ulem}

\newcommand{\OX}{\Omega_{\rm X}}
\newcommand{\LCDM}{\Lambda\mathrm{CDM}}
\begin{document}
\title{Does Dark Energy Evolve \emph{and} Cluster?}
\author{Maria Luisa Gallavotti}
\affiliation{Instituut-Lorentz, Leiden University, PO Box 9506, 2300 RA Leiden, The Netherlands}
\author{Alessandra Silvestri}
\affiliation{Instituut-Lorentz, Leiden University, PO Box 9506, 2300 RA Leiden, The Netherlands}
\date{\today}
\begin{abstract}
Recent cosmological observations have renewed interest in dynamical dark energy, and naturally raise the question of whether departures from the standard expansion history are accompanied by modifications of gravitational dynamics. We address this question through a joint, non-parametric reconstruction of the effective dark-energy contribution to the expansion and the gravitational responses governing clustering and lensing, complemented by a combined covariance-mode analysis, without imposing model-specific relations between the sectors. Combining CMB, DESI DR2 BAO, and Pantheon+ supernova observations, and subsequently adding large scale structure observations including DES Year 3 $6\times2$pt data and other surveys, we recover a broad maximum in the effective dark-energy density near $z\simeq0.6$--$0.7$. The gravitational responses exhibit suggestive low-redshift enhancements over their GR values, approximately coincident with the background feature. Adding large scale structure information preserves the background feature but favours a mildly reduced lensing response. The combined covariance-mode analysis identifies the combinations of background and gravitational responses most tightly constrained by current observations and quantifies their departures from $\Lambda$CDM. The leading modes are decisively background dominated, while the gravitational sector remains less precisely constrained. Thus, present data cannot yet determine whether the suggestive low-redshift features in the two sectors share a common physical origin, a question that ongoing Stage-IV surveys will be able to address.

\end{abstract}
\maketitle

\emph{Introduction.---}
The nature of dark energy (DE), responsible for the
accelerated expansion of the Universe, remains one of
the fundamental open questions in physics. Recent
measurements by the Dark Energy Spectroscopic Instrument
(DESI), combined with other cosmological probes, have
revived interest in an evolving dark energy sector
\cite{DESI:2024mwx,DESI:2025fii,DES:2026jmi}. If confirmed at sufficient
statistical significance, such evolution would rule out
a cosmological constant, $\Lambda$, as the sole driver
of cosmic acceleration. Yet a departure from the
expansion history predicted by the standard cosmological
model, $\LCDM$, does not uniquely identify the underlying
physics. In particular, an evolving dark energy component,
especially one crossing the phantom divide, may require
a richer theoretical framework than canonical
quintessence. Many viable realizations involve additional
degrees of freedom or non-minimal gravitational interactions~\cite{Ye:2024ywg,Reboucas:2026ljj,Wolf:2024stt},
potentially leaving correlated signatures in the growth
of cosmic structure and gravitational lensing.

This motivates a simple question: \emph{does dark energy
evolve and cluster?} Here, we use DE in a broad sense,
encompassing both additional dynamical fields and
modifications of gravity that can drive cosmic acceleration.
We investigate whether departures from the standard
expansion history are accompanied by modifications of the
gravitational response to inhomogeneities, thereby probing
the background and perturbation sectors jointly.

We address it by jointly reconstructing the background expansion, matter-clustering response, and gravitational-lensing response from the latest data, without fixing a theory-dependent relation among them. We then identify which \emph{combined} combinations are constrained by the likelihood and how far their posterior means lie from the $\LCDM$ reference. As we will discuss, the hierarchy of constrained modes reflects the information content of present datasets and need not coincide with the hierarchy of underlying physical departures. The full reconstruction, individual-sector analyses, dataset dependence, and scalar--tensor interpretation are developed in a companion article \cite{CompanionPRD}.

\emph{Framework and data.---}
On large scales the  universe is homogeneous and isotropic and is well-described by the flat Friedmann Lemaitre- Robertson-Walker (FLRW) metric with line element
$ds^2=-dt^2+a(t)^2d\bar{x}^2$. The scale factor $a(t)$ describes the background expansion and its evolution is determined by the Friedmann equation
\begin{equation}
 H^2(a)=H_0^2\left[\Omega_r a^{-4}+\Omega_m a^{-3}+\OX(a)\right],
 \label{eq:background}
\end{equation}
where $\OX(a)$ encompasses any contribution to the Friedmann equation other than radiation and matter. 
We can  parametrize the linear gravitational dynamics through
\begin{align}
 k^2\Psi&=-\frac{a^2}{2M_P^2}\,\mu(a,k)\rho_m\Delta_m,\\
 k^2(\Phi+\Psi)&=-\frac{a^2}{M_P^2}\,\Sigma(a,k)\rho_m\Delta_m.
 \label{eq:response}
\end{align}
Here $\mu$ controls the response of the Newtonian potential, $\Psi$, to matter perturbations, $\Delta_m$. $\Sigma$ controls the Weyl potential, $\Phi+\Psi$, describing relativistic geodesics, and is therefore constrained by weak gravitational lensing of galaxies as well as the CMB. The reference model, $\Lambda$CDM, has $\OX(a)=\Omega_\Lambda$ and $\mu=\Sigma=1$. In this work we will focus on the time-dependence of $\mu$ and $\Sigma$, which is the dependence to which current large-scale-structure (LSS) data are most sensitive. The effective equation of state inferred from the expansion is $w_{\rm X}=-1-\frac13\,d\ln\OX/d\ln a$; in general it need not describe a separately conserved fluid, as it can simply correspond to an effective fluid encoding the modifications of gravity to the Friedmann equation.

Our reconstruction is based on a non-parametric method, where we specify each function at 11 scale-factor nodes spanning the redshift interval $0\le z\le4$, and use a  cubic spline interpolation to obtain a continuous function. We then evolve the full system of equations for the background and linear perturbations with \texttt{MGCAMB}~\cite{Wang:2023tjj}. The present-day $\OX$ node is fixed by the flatness condition. We consider reconstructions both with and without a theory correlation prior, derived from an ensemble of stable Horndeski gravity models~\cite{Espejo:2018hxa, Pogosian:2021mcs, Raveri:2021dbu}. The latter acts as a smoothing filter which regularizes poorly constrained temporal structure, while introducing a mild correlation in the reconstructed sectors. For our analysis we focus on the best constrained modes for which the posterior information dominates over the prior one.

We consider two combinations of data sets:  \emph{Baseline} which combines Planck 2018 low-$\ell$~\cite{Planck:2019nip,Pagano:2019tci} and ACT DR6~\cite{AtacamaCosmologyTelescope:2025blo} primary-CMB measurements, Planck PR4 CMB lensing~\cite{Carron:2022eyg}, DESI DR2  BAO~\cite{DESI:2025zpo,DESI:2025zgx}, and Pantheon+ supernovae~\cite{Scolnic:2021amr,Brout:2022vxf}; and  \emph{Baseline+LSS} adds DES Year 3 (Y3) galaxy clustering, cosmic shear, galaxy--galaxy lensing~\cite{DES:2021wwk,DES:2021bpo} and cross-correlations with CMB lensing from Planck and SPT~\cite{DES:2022qdz,DES:2022xxr,DES:2022urg}, together with SDSS DR16 BAO+RSD~\cite{Alam:2020sor}. The latter uses the combined ACT DR6+Planck+SPT-3G CMB-lensing likelihood in place of Planck lensing alone~\cite{ACT:2025qjh,ACT:2023kun,ACT:2023dou,ACT:2023ubw,SPT-3G:2024atg}. We apply published conservative scale cuts~\cite{DES:2022ccp,DES:2022qdz,DES:2022xxr} and model intrinsic alignments with TATT~\cite{Blazek:2017wbz}. We introduce the prior via the joint covariance $\mathbf{C}_{\rm H}$ derived from the theory ensemble~\cite{Espejo:2018hxa} as an additional Gaussian prior, adopting the floating-fiducial prescription of Ref.~\cite{Crittenden:2011aa}. Full likelihood and implementation details, including the prior construction and intrinsic-alignment tests, are given in the companion article~\cite{CompanionPRD}. 

\begin{figure}[t]
\centering
\includegraphics[width=0.9\columnwidth]{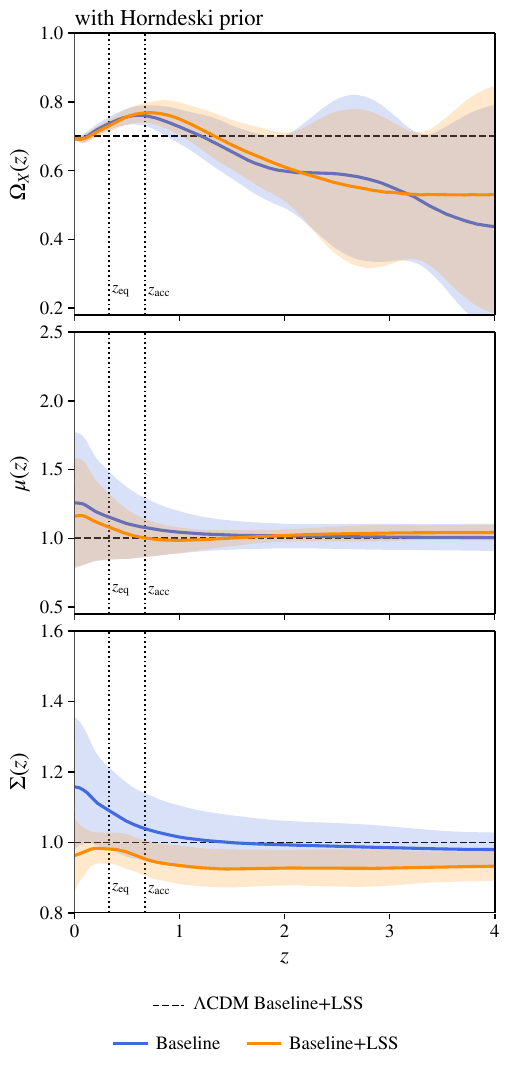}
\caption{Joint reconstruction of $\OX$, $\mu$ and $\Sigma$ for Baseline and Baseline+LSS, with the Horndeski correlation prior. Curves show posterior medians and bands pointwise 68\% credible intervals.}
\label{fig:reconstruction}
\end{figure}

\emph{The reconstructed histories.---}
Figure~\ref{fig:reconstruction} summarizes the three reconstructed functions. The most prominent feature is a broad maximum of $\OX$ near $z\simeq0.6$--$0.7$. Its location remains similar upon adding LSS information.  In~\cite{CompanionPRD} we carry out an auxiliary reconstruction with $\mu=\Sigma=1$, showing that  the background turnover is already present in CMB+DESI BAO, with supernova and LSS data refining its location and shape. The maximum of $\OX$ corresponds to an effective phantom-divide crossing, with $w_{\rm X}>-1$ below the maximum and $w_{\rm X}<-1$ above it, as redshift increases, wherever $\OX>0$.

The perturbation sector exhibits milder, less significant features, which however must be interpreted in light of its weaker observational constraints. In Baseline, both $\mu$ and $\Sigma$ tend above unity over part of the low-redshift range, approximately coinciding with the rise of $\OX$ above $\Omega_\Lambda$. The inclusion of LSS information reduces their allowed freedom and leaves a mild low-redshift enhancement in $\mu$, while shifting $\Sigma$ below unity. This trend suggests a reduced effective lensing response and persists when TATT is replaced by NLA within the tested modeling choices, see accompanying paper~\cite{CompanionPRD}. In~\cite{CompanionPRD} we also show that the corresponding inferred gravitational slip, $\gamma=2\Sigma/\mu-1$, tends below unity, while remaining broadly compatible with GR, and offer an interpretation of these findings within the Horndeski gravity framework.

Our central question for this paper is whether current observations provide evidence for correlated departures from $\Lambda$CDM in the background and perturbation sectors. To address this, we now turn to the combined modes most strongly constrained by the data relative to the prior, constructed from the nodes of all three reconstructed functions.

\emph{Which departures are measured?---}
A high-dimensional reconstruction can contain apparent features in poorly constrained directions. To distinguish information provided by the data from
prior regularization, we perform a covariant principal-component analysis (CPCA). On the 30-dimensional subspace constrained by the floating-fiducial smoothing prior, we solve
\begin{equation}
 C_\Pi\bm\Psi_\alpha=R_\alpha C_p\bm\Psi_\alpha,
 \label{eq:cpca}
\end{equation}
where $C_\Pi$ and $C_p$ are the reference-prior and posterior covariances. $\bm\Psi_\alpha$ is the $\alpha$th principal mode and $R_\alpha$ measures the reduction in its variance relative to the prior. A large $R_\alpha$ indicates substantial variance reduction beyond the prior. To assess whether a constrained mode also exhibits
a departure from the reference model, we evaluate
its displacement via the signal-to-noise (SNR) ratio
\begin{equation}
{\rm SNR}_\alpha=\frac{\left|\bm\Psi_\alpha^T(\bar{\bm f}-\bm f_{0})\right|}
 {\sqrt{\bm\Psi_\alpha^T C_p\bm\Psi_\alpha}}\,,
 \label{eq:displacement}
\end{equation}
where $\bm f$ collects the node values of the three reconstructed functions, $\bar{\bm f}$ is its posterior mean, and $\bm f_{0}$ is the corresponding $\Lambda$CDM reference vector.
Thus, the eigenvalues measure the constraining power of the data, while the displacements quantify departures from the reference model along individual modes.

Here, we focus on the \emph{combined} modes of the three
functions to quantify \emph{any} departure from $\Lambda$CDM along these directions, and to identify the  combinations
of background and perturbation variables to which current
data are most sensitive. In practice this means that we perform CPCA on the full vector $\bm f$.  Unlike
separate-sector analyses, these modes can contain
contributions from all three functions, and their
number need not equal the sum of the individually
constrained modes \cite{CompanionPRD}.

To characterize their composition, we define the
relative contributions from $\OX$, $\mu$, and
$\Sigma$, weighted by the corresponding blocks
of the reference-prior covariance:
\begin{equation}
   \mathcal F _{a,s} =
    \frac{\boldsymbol\Psi_{a,s}^{\rm T}
    \mathbf{C}_{\Pi,ss}\boldsymbol\Psi_{a,s}}
    {\sum_t \boldsymbol\Psi_{a,t}^{\rm T}
    \mathbf{C}_{\Pi,tt}\boldsymbol\Psi_{a,t}},
   \label{eq:mode_composition}
\end{equation}
where $s,t\in\{\Omega_{\rm X},\mu,\Sigma\}$.
Here $\boldsymbol\Psi_{a,s}$ contains the weights of the eleven nodes belonging to sector $s$, and $\mathbf{C}_{\Pi,ss}$ is the corresponding block of the reference-prior covariance. The resulting fractions sum to unity and characterize the relative contributions of the three functions to each mode, accounting for their different prior covariance scales. Because the prior also has cross-sector correlations, these fractions are a useful measure of mode composition, not a unique decomposition of the total variance. 
\begin{figure}
    \centering
\includegraphics[width=\columnwidth]{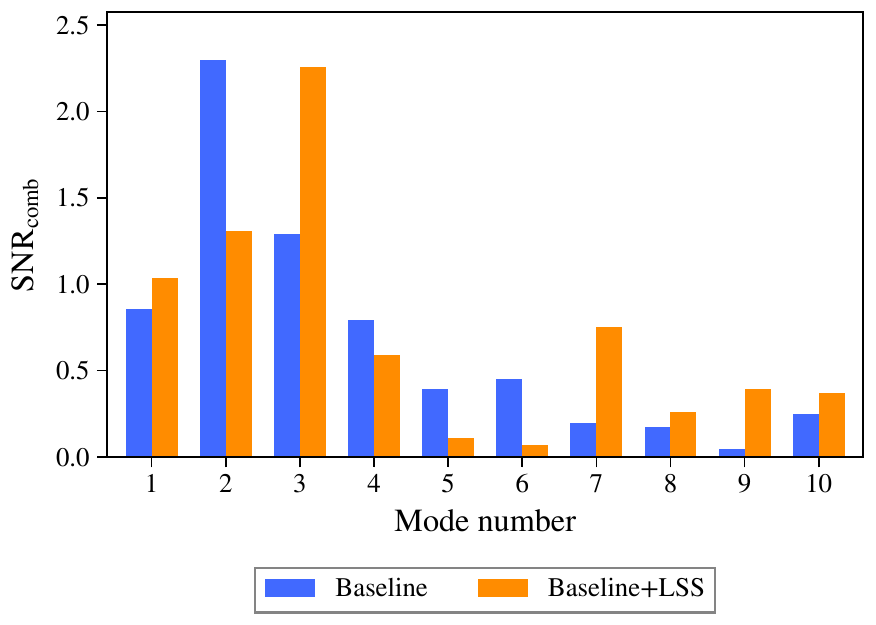}
    \caption{Displacements from $\Lambda$CDM of the combined CPCA modes, ordered by measurement precision, $R_\alpha$~\eqref{eq:cpca}.}
\label{fig:displacements}
\end{figure}

\begin{figure}
    \centering
\includegraphics[width=0.9\columnwidth]{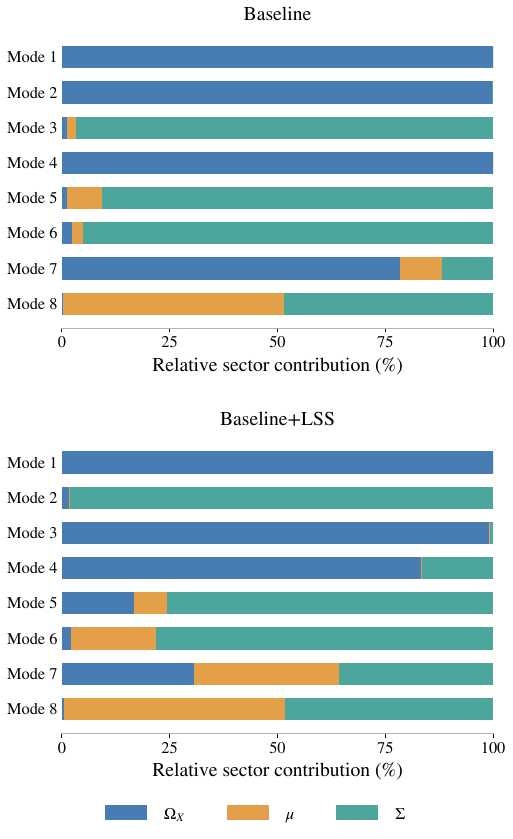}
    \caption{ 
    Relative sector contributions to the best constrained combined CPCA modes for the joint reconstructions with Baseline (top panel) and Baseline+LSS (bottom panel). Contributions provide a measure of mode composition, and are defined by Eq.~\eqref{eq:mode_composition}.
}
\label{fig:joint_cpca_composition}
\end{figure}

In Fig.~\ref{fig:displacements} we show the displacement of the combined modes for both the Baseline and Baseline+LSS data combinations. In Baseline, the most displaced combined mode is the second best measured one, i.e. mode 2, with ${\rm SNR}_2\simeq2.3$, whereas in Baseline+LSS it is mode 3, again with ${\rm SNR}_3\simeq2.3$. Both of these modes are strongly dominated by $\OX$. In Fig.~\ref{fig:joint_cpca_composition} we show the composition of the first $8$ combined modes, i.e. the relative contributions from the three functions $\OX$, $\mu$ and $\Sigma$. The first mode in both combinations is entirely or almost entirely background dominated, although its displacement is only of order unity. By contrast, the third Baseline mode and second Baseline+LSS mode are mostly sensitive to $\Sigma$, with smaller displacements of ${\rm SNR}\simeq1.3$ in both cases. Mixed background--lensing directions also appear, such as the fourth mode of Baseline+LSS, whose displacement remains below unity.

The dominance of $\OX$ in the leading combined modes is expected given the strong constraining power of geometric data on the background expansion. It should therefore \emph{not} be interpreted as evidence that the underlying physics acts only on the background, or that the lower-significance trends in $\mu$ and $\Sigma$ are physically independent of the turnover in $\OX$. The perturbation sector is constrained less precisely, so correlated departures could remain unresolved or project onto lower-ranked modes. As such, the present reconstruction does not establish such correlations. 

\emph{Physical implications.---} The joint reconstruction reveals an intriguing picture. The low-redshift maximum in $\Omega_{\rm X}(z)$, corresponding to a crossing of the effective phantom divide, persists when modifications to gravitational clustering and lensing are allowed and remains robust to the inclusion of LSS information. Interestingly, the Baseline reconstruction also exhibits enhancements
of approximately $20\%$ in both $\mu$ and $\Sigma$ over a similar low-redshift range, with both functions approaching their GR values at higher redshifts. Although suggestive of a possible common physical origin, these departures remain statistically weak, given the limited constraining power of current observations on the perturbation sector.
\begin{figure}
    \includegraphics[width=\columnwidth]{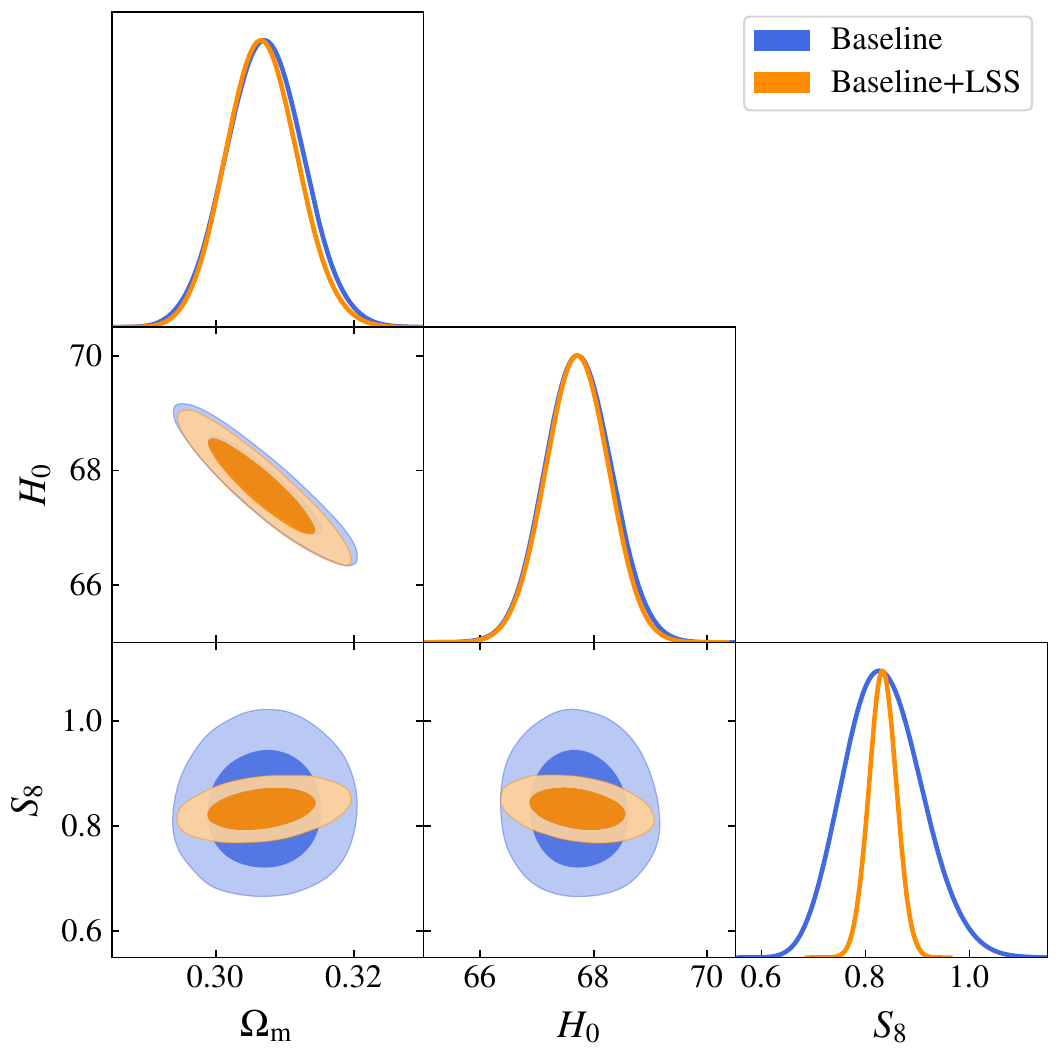}
    \caption{The 68\% and 95\% credible contours for $\Omega_{\rm m}$, $H_0$, and $S_8$, from the Baseline and Baseline+LSS data with the Horndeski prior.
    }
    \label{fig:triangle_joint_prior}
\end{figure} 
The inclusion of LSS information leaves $\mu$ largely unchanged but shifts $\Sigma$ below unity, favouring a reduced effective lensing response. This preference can be accommodated without a corresponding reduction in the inferred matter fluctuation amplitude $S_8$, as illustrated by the stability of the cosmological parameter constraints in Fig.~\ref{fig:triangle_joint_prior}.

A scalar--tensor interpretation can accommodate competing modifications to the effective gravitational coupling and a scalar-mediated fifth force, but the reconstructed functions do not by themselves establish a viable theory. We present a detailed generalized Brans--Dicke mapping, as well as the intrinsic-alignment and prior-dependence tests in the companion article \cite{CompanionPRD}.

Upcoming LSS measurements from the ongoing Stage-IV surveys can change the CPCA hierarchy by constraining gravity-sensitive directions more strongly. If the low-redshift trends in $\mu$ and $\Sigma$ persist, their displacements may become more decisive. 

\emph{Conclusions.---}
We have investigated whether indications of evolving dark energy are accompanied by modifications of linear gravitational dynamics. A joint non-parametric reconstruction finds a persistent low-redshift maximum in the effective DE contribution while allowing independent changes in growth and lensing. The reconstructed gravitational responses show milder, dataset-dependent features, including approximately coincident low-redshift enhancements in the Baseline analysis and a preference for reduced lensing response when LSS information is added. Combined CPCA identifies the directions currently best measured, which are predominantly background dominated.  This reflects the relative observational precision and does not rule out a common origin for background and perturbation departures. Present data therefore provide a suggestive but inconclusive answer to whether evolving dark energy also modifies gravitational clustering. Future precision LSS observations will test whether the gravitational features persist and become statistically resolved. The complete reconstruction, theoretical interpretation, and observational robustness tests are presented in our companion article \cite{CompanionPRD}.

\begin{acknowledgments}
We thank Matteo Martinelli, Gen Ye, Hamid Mirpoorian, and Marco Raveri for discussions. This work was supported by the European Research Council under Grant No.~101126217 (GraviPULSE) and by SURF Cooperative grant EINF-18786.
\end{acknowledgments}

%%%%%%%%%%%%%%%%%%%%%%%%%%%%%%%%%%%%%%%%%%%%%%%%%%
% Bibliography
%%%%%%%%%%%%%%%%%%%%%%%%%%%%%%%%%%%%%%%%%%%%%%%%%%

\bibliographystyle{apsrev4-2}
\bibliography{references}

@article{DESI:2025fii,
    author = "Lodha, K. and others",
    collaboration = "DESI",
    title = "{Extended dark energy analysis using DESI DR2 BAO measurements}",
    eprint = "2503.14743",
    archivePrefix = "arXiv",
    primaryClass = "astro-ph.CO",
    reportNumber = "FERMILAB-PUB-25-0164-PPD",
    doi = "10.1103/w4c6-1r5j",
    journal = "Phys. Rev. D",
    volume = "112",
    number = "8",
    pages = "083511",
    year = "2025"
}

@article{Wolf:2024stt,
    author = "Wolf, William J. and Ferreira, Pedro G. and Garc{\'\i}a-Garc{\'\i}a, Carlos",
    title = "{Matching current observational constraints with nonminimally coupled dark energy}",
    eprint = "2409.17019",
    archivePrefix = "arXiv",
    primaryClass = "astro-ph.CO",
    doi = "10.1103/PhysRevD.111.L041303",
    journal = "Phys. Rev. D",
    volume = "111",
    number = "4",
    pages = "L041303",
    year = "2025"
}

@article{Ye:2024ywg,
    author = "Ye, Gen and Martinelli, Matteo and Hu, Bin and Silvestri, Alessandra",
    title = "{Hints of Nonminimally Coupled Gravity in DESI 2024 Baryon Acoustic Oscillation Measurements}",
    eprint = "2407.15832",
    archivePrefix = "arXiv",
    primaryClass = "astro-ph.CO",
    doi = "10.1103/PhysRevLett.134.181002",
    journal = "Phys. Rev. Lett.",
    volume = "134",
    number = "18",
    pages = "181002",
    year = "2025"
}

@article{DES:2026jmi,
    journal = {arXiv e-prints},
    author = "Abbott, T. M. C. and others",
    collaboration = "DES",
    title = "{Constraints on Dynamical Dark Energy from Multiple Probes in the Full Dark Energy Survey}",
    eprint = "2605.27221",
    archivePrefix = "arXiv",
    primaryClass = "astro-ph.CO",
    reportNumber = "DES-2026-0979, FERMILAB-PUB-26-0306-PPD",
    month = "5",
    year = "2026"
}

@article{Reboucas:2026ljj,
    journal = {arXiv e-prints},
    author = "Rebou{\c{c}}as, Jo{\~a}o and Brando, Guilherme and Falciano, Felipe T. and Miranda, Vivian",
    title = "{The sound of dynamical dark energy and modified gravity}",
    eprint = "2606.00411",
    archivePrefix = "arXiv",
    primaryClass = "astro-ph.CO",
    month = "5",
    year = "2026"
}

@article{DESI:2024mwx,
    author = "Adame, A. G. and others",
    collaboration = "DESI",
    title = "{DESI 2024 VI: cosmological constraints from the measurements of baryon acoustic oscillations}",
    eprint = "2404.03002",
    archivePrefix = "arXiv",
    primaryClass = "astro-ph.CO",
    reportNumber = "FERMILAB-PUB-24-0154-PPD",
    doi = "10.1088/1475-7516/2025/02/021",
    journal = "JCAP",
    volume = "02",
    pages = "021",
    year = "2025"
}

@article{Scolnic:2021amr,
    author = "Scolnic, Dan and others",
    title = "{The Pantheon+ Analysis: The Full Data Set and Light-curve Release}",
    eprint = "2112.03863",
    archivePrefix = "arXiv",
    primaryClass = "astro-ph.CO",
    doi = "10.3847/1538-4357/ac8b7a",
    journal = "Astrophys. J.",
    volume = "938",
    number = "2",
    pages = "113",
    year = "2022"
}

@article{Planck:2019nip,
    author = "Aghanim, N. and others",
    collaboration = "Planck",
    title = "{Planck 2018 results. V. CMB power spectra and likelihoods}",
    eprint = "1907.12875",
    archivePrefix = "arXiv",
    primaryClass = "astro-ph.CO",
    doi = "10.1051/0004-6361/201936386",
    journal = "Astron. Astrophys.",
    volume = "641",
    pages = "A5",
    year = "2020"
}

@article{Carron:2022eyg,
    author = "Carron, Julien and Mirmelstein, Mark and Lewis, Antony",
    title = "{CMB lensing from Planck PR4~maps}",
    eprint = "2206.07773",
    archivePrefix = "arXiv",
    primaryClass = "astro-ph.CO",
    doi = "10.1088/1475-7516/2022/09/039",
    journal = "JCAP",
    volume = "09",
    pages = "039",
    year = "2022"
}

@article{Espejo:2018hxa,
    author = "Espejo, Juan and Peirone, Simone and Raveri, Marco and Koyama, Kazuya and Pogosian, Levon and Silvestri, Alessandra",
    title = "{Phenomenology of Large Scale Structure in scalar-tensor theories: joint prior covariance of $w_{\textrm{DE}}$, $\Sigma$ and $\mu$ in Horndeski}",
    eprint = "1809.01121",
    archivePrefix = "arXiv",
    primaryClass = "astro-ph.CO",
    doi = "10.1103/PhysRevD.99.023512",
    journal = "Phys. Rev. D",
    volume = "99",
    number = "2",
    pages = "023512",
    year = "2019"
}

@article{Raveri:2021dbu,
    author = "Raveri, Marco and Pogosian, Levon and Martinelli, Matteo and Koyama, Kazuya and Silvestri, Alessandra and Zhao, Gong-Bo",
    title = "{Principal reconstructed modes of dark energy and gravity}",
    eprint = "2107.12990",
    archivePrefix = "arXiv",
    primaryClass = "astro-ph.CO",
    doi = "10.1088/1475-7516/2023/02/061",
    journal = "JCAP",
    volume = "02",
    pages = "061",
    year = "2023"
}

@article{Pogosian:2021mcs,
    author = "Pogosian, Levon and Raveri, Marco and Koyama, Kazuya and Martinelli, Matteo and Silvestri, Alessandra and Zhao, Gong-Bo and Li, Jian and Peirone, Simone and Zucca, Alex",
    title = "{Imprints of cosmological tensions in reconstructed gravity}",
    eprint = "2107.12992",
    archivePrefix = "arXiv",
    primaryClass = "astro-ph.CO",
    doi = "10.1038/s41550-022-01808-7",
    journal = "Nature Astron.",
    volume = "6",
    number = "12",
    pages = "1484--1490",
    year = "2022"
}

@article{Wang:2023tjj,
    author = "Wang, Zhuangfei and Mirpoorian, Seyed Hamidreza and Pogosian, Levon and Silvestri, Alessandra and Zhao, Gong-Bo",
    title = "{New MGCAMB tests of gravity with CosmoMC and Cobaya}",
    eprint = "2305.05667",
    archivePrefix = "arXiv",
    primaryClass = "astro-ph.CO",
    doi = "10.1088/1475-7516/2023/08/038",
    journal = "JCAP",
    volume = "08",
    pages = "038",
    year = "2023"
}

@article{Crittenden:2011aa,
    author = "Crittenden, Robert G. and Zhao, Gong-Bo and Pogosian, Levon and Samushia, Lado and Zhang, Xinmin",
    title = "{Fables of reconstruction: controlling bias in the dark energy equation of state}",
    eprint = "1112.1693",
    archivePrefix = "arXiv",
    primaryClass = "astro-ph.CO",
    doi = "10.1088/1475-7516/2012/02/048",
    journal = "JCAP",
    volume = "02",
    pages = "048",
    year = "2012"
}

@article{Pagano:2019tci,
    author = "Pagano, L. and Delouis, J. -M. and Mottet, S. and Puget, J. -L. and Vibert, L.",
    title = "{Reionization optical depth determination from Planck HFI data with ten percent accuracy}",
    eprint = "1908.09856",
    archivePrefix = "arXiv",
    primaryClass = "astro-ph.CO",
    doi = "10.1051/0004-6361/201936630",
    journal = "Astron. Astrophys.",
    volume = "635",
    pages = "A99",
    year = "2020"
}

@article{AtacamaCosmologyTelescope:2025blo,
    author = "Louis, Thibaut and others",
    collaboration = "Atacama Cosmology Telescope",
    title = "{The Atacama Cosmology Telescope: DR6 power spectra, likelihoods and {\ensuremath{\Lambda}}CDM parameters}",
    eprint = "2503.14452",
    archivePrefix = "arXiv",
    primaryClass = "astro-ph.CO",
    reportNumber = "FERMILAB-PUB-25-0071-PPD",
    doi = "10.1088/1475-7516/2025/11/062",
    journal = "JCAP",
    volume = "11",
    pages = "062",
    year = "2025"
}

@article{DESI:2025zpo,
    journal = {arXiv e-prints},
    author = "Abdul Karim, M. and others",
    collaboration = "DESI",
    title = "{DESI DR2 Results I: Baryon Acoustic Oscillations from the Lyman Alpha Forest}",
    eprint = "2503.14739",
    archivePrefix = "arXiv",
    primaryClass = "astro-ph.CO",
    month = "3",
    year = "2025"
}

@article{DESI:2025zgx,
    journal = {arXiv e-prints},
    author = "Abdul Karim, M. and others",
    collaboration = "DESI",
    title = "{DESI DR2 Results II: Measurements of Baryon Acoustic Oscillations and Cosmological Constraints}",
    eprint = "2503.14738",
    archivePrefix = "arXiv",
    primaryClass = "astro-ph.CO",
    month = "3",
    year = "2025"
}

@article{Brout:2022vxf,
    author = "Brout, Dillon and others",
    title = "{The Pantheon+ Analysis: Cosmological Constraints}",
    eprint = "2202.04077",
    archivePrefix = "arXiv",
    primaryClass = "astro-ph.CO",
    doi = "10.3847/1538-4357/ac8e04",
    journal = "Astrophys. J.",
    volume = "938",
    number = "2",
    pages = "110",
    year = "2022"
}

@article{Alam:2020sor,
    author = "Alam, Shadab and others",
    collaboration = "eBOSS",
    title = "{Completed SDSS-IV extended Baryon Oscillation Spectroscopic Survey:
    Cosmological implications from two decades of spectroscopic surveys at the Apache
    Point Observatory}",
    eprint = "2007.08991",
    archivePrefix = "arXiv",
    primaryClass = "astro-ph.CO",
    doi = "10.1103/PhysRevD.103.083533",
    journal = "Phys. Rev. D",
    volume = "103",
    number = "8",
    pages = "083533",
    year = "2021"
}

@article{ACT:2025qjh,
    author = "Qu, Frank J. and others",
    collaboration = "ACT, SPT-3G",
    title = "{Unified and Consistent Structure Growth Measurements from Joint ACT, SPT, and Planck CMB Lensing}",
    eprint = "2504.20038",
    archivePrefix = "arXiv",
    primaryClass = "astro-ph.CO",
    reportNumber = "FERMILAB-PUB-25-0279-PPD",
    doi = "10.1103/k5yr-3h6d",
    journal = "Phys. Rev. Lett.",
    volume = "136",
    number = "2",
    pages = "021001",
    year = "2026"
}

@article{SPT-3G:2024atg,
    author = "Ge, F. and others",
    collaboration = "SPT-3G",
    title = "{Cosmology from CMB lensing and delensed EE power spectra using 2019{\textendash}2020 SPT-3G polarization data}",
    eprint = "2411.06000",
    archivePrefix = "arXiv",
    primaryClass = "astro-ph.CO",
    reportNumber = "FERMILAB-PUB-24-0840-PPD",
    doi = "10.1103/PhysRevD.111.083534",
    journal = "Phys. Rev. D",
    volume = "111",
    number = "8",
    pages = "083534",
    year = "2025"
}

@article{ACT:2023kun,
    author = "Madhavacheril, Mathew S. and others",
    collaboration = "ACT",
    title = "{The Atacama Cosmology Telescope: DR6 Gravitational Lensing Map and Cosmological Parameters}",
    eprint = "2304.05203",
    archivePrefix = "arXiv",
    primaryClass = "astro-ph.CO",
    reportNumber = "FERMILAB-PUB-23-206-PPD",
    doi = "10.3847/1538-4357/acff5f",
    journal = "Astrophys. J.",
    volume = "962",
    number = "2",
    pages = "113",
    year = "2024"
}

@article{ACT:2023dou,
    author = "Qu, Frank J. and others",
    collaboration = "ACT",
    title = "{The Atacama Cosmology Telescope: A Measurement of the DR6 CMB Lensing Power Spectrum and Its Implications for Structure Growth}",
    eprint = "2304.05202",
    archivePrefix = "arXiv",
    primaryClass = "astro-ph.CO",
    reportNumber = "FERMILAB-PUB-23-237-PPD, FERMILAB-PUB-23-237-PPD",
    doi = "10.3847/1538-4357/acfe06",
    journal = "Astrophys. J.",
    volume = "962",
    number = "2",
    pages = "112",
    year = "2024"
}

@article{ACT:2023ubw,
    author = "MacCrann, Niall and others",
    collaboration = "ACT",
    title = "{The Atacama Cosmology Telescope: Mitigating the Impact of Extragalactic Foregrounds for the DR6 Cosmic Microwave Background Lensing Analysis}",
    eprint = "2304.05196",
    archivePrefix = "arXiv",
    primaryClass = "astro-ph.CO",
    reportNumber = "FERMILAB-PUB-23-236-PPD",
    doi = "10.3847/1538-4357/ad2610",
    journal = "Astrophys. J.",
    volume = "966",
    number = "1",
    pages = "138",
    year = "2024"
}

@article{DES:2021wwk,
    author = "Abbott, T. M. C. and others",
    collaboration = "DES",
    title = "{Dark Energy Survey Year 3 results: Cosmological constraints from galaxy clustering and weak lensing}",
    eprint = "2105.13549",
    archivePrefix = "arXiv",
    primaryClass = "astro-ph.CO",
    reportNumber = "FERMILAB-PUB-21-221-AE, DES-2020-0617",
    doi = "10.1103/PhysRevD.105.023520",
    journal = "Phys. Rev. D",
    volume = "105",
    number = "2",
    pages = "023520",
    year = "2022"
}

@article{DES:2021bpo,
    author = "Porredon, A. and others",
    collaboration = "DES",
    title = "{Dark Energy Survey Year 3 results: Cosmological constraints from galaxy clustering and galaxy-galaxy lensing using the MagLim lens sample}",
    eprint = "2105.13546",
    archivePrefix = "arXiv",
    primaryClass = "astro-ph.CO",
    reportNumber = "FERMILAB-PUB-21-251-AE",
    doi = "10.1103/PhysRevD.106.103530",
    journal = "Phys. Rev. D",
    volume = "106",
    number = "10",
    pages = "103530",
    year = "2022"
}

@article{DES:2022qdz,
    author = "Omori, Y. and others",
    collaboration = "DES, SPT",
    title = "{Joint analysis of Dark Energy Survey Year 3 data and CMB lensing from SPT and Planck. I. Construction of CMB lensing maps and modeling choices}",
    eprint = "2203.12439",
    archivePrefix = "arXiv",
    primaryClass = "astro-ph.CO",
    reportNumber = "FERMILAB-PUB-22-194-PPD",
    doi = "10.1103/PhysRevD.107.023529",
    journal = "Phys. Rev. D",
    volume = "107",
    number = "2",
    pages = "023529",
    year = "2023"
}

@article{DES:2022xxr,
    author = "Chang, C. and others",
    collaboration = "DES, SPT",
    title = "{Joint analysis of Dark Energy Survey Year 3 data and CMB lensing from SPT and Planck. II. Cross-correlation measurements and cosmological constraints}",
    eprint = "2203.12440",
    archivePrefix = "arXiv",
    primaryClass = "astro-ph.CO",
    reportNumber = "FERMILAB-PUB-22-098-PPD",
    doi = "10.1103/PhysRevD.107.023530",
    journal = "Phys. Rev. D",
    volume = "107",
    number = "2",
    pages = "023530",
    year = "2023"
}

@article{DES:2022urg,
    author = "Abbott, T. M. C. and others",
    collaboration = "DES, SPT",
    title = "{Joint analysis of Dark Energy Survey Year 3 data and CMB lensing from SPT and Planck. III. Combined cosmological constraints}",
    eprint = "2206.10824",
    archivePrefix = "arXiv",
    primaryClass = "astro-ph.CO",
    reportNumber = "FERMILAB-PUB-22-475-PPD",
    doi = "10.1103/PhysRevD.107.023531",
    journal = "Phys. Rev. D",
    volume = "107",
    number = "2",
    pages = "023531",
    year = "2023"
}

@article{DES:2022ccp,
    author = "Abbott, T. M. C. and others",
    collaboration = "DES",
    title = "{Dark Energy Survey Year 3 results: Constraints on extensions to {\ensuremath{\Lambda}}CDM with weak lensing and galaxy clustering}",
    eprint = "2207.05766",
    archivePrefix = "arXiv",
    primaryClass = "astro-ph.CO",
    reportNumber = "FERMILAB-PUB-22-470-PPD",
    doi = "10.1103/PhysRevD.107.083504",
    journal = "Phys. Rev. D",
    volume = "107",
    number = "8",
    pages = "083504",
    year = "2023"
}

@article{Blazek:2017wbz,
    author = "Blazek, Jonathan and MacCrann, Niall and Troxel, M. A. and Fang, Xiao",
    title = "{Beyond linear galaxy alignments}",
    eprint = "1708.09247",
    archivePrefix = "arXiv",
    primaryClass = "astro-ph.CO",
    doi = "10.1103/PhysRevD.100.103506",
    journal = "Phys. Rev. D",
    volume = "100",
    number = "10",
    pages = "103506",
    year = "2019"
}

@misc{CompanionPRD,
  author = {Gallavotti, Maria Luisa and Silvestri, Alessandra},
  title = {{Reconstruction of Dynamical Dark Energy and Gravity
            from latest cosmological data}},
  note = {Companion paper, submitted to arXiv},
  year = {2026}
}

\end{document}